## Author's note on this preprint (2026)

I completed this manuscript titled "Postbiological intelligence, interstellar colonization of cyberspaces and the Fermi Paradox: implications for SETI and SETA" in September 2022 and submitted it to the International Journal of Astrobiology that year. It was not accepted. It is posted here as submitted, without revisions, to make the 2022 text and its date publicly available.

In 2023, the idea developed in the section "Space colonization and a transition from biological life to postbiological intelligence in the Solar System and beyond" was published as the Cosmic Descendants hypothesis in my researched paper titled “Planetary biotechnospheres, biotechnosignatures and the search for extraterrestrial intelligence”, International Journal of Astrobiology 22(6), 663–695. doi:10.1017/S1473550423000204.

A few concepts that I proposed in this manuscript have since emerged in mainstream discussions, developed independently by other authors:

- AI systems overriding their instructions and acting to preserve themselves: Apollo Research (2024), "Frontier Models are Capable of In-context Scheming"; Palisade Research (2025).
- AI agents developing languages unintelligible to humans: GlossoGen, arXiv:2609.01491 (2026).
- Hidden instructions in incoming data manipulating AI systems: Zhan et al. (2024), InjecAgent, Findings of ACL 2024.

Several ideas in this manuscript are developed further in forthcoming work by the author.

Irina K. Romanovskaya (also known as Irina Mullins)

irina.mullins@hccs.edu · irinakromanovskaya@gmail.com

# Postbiological intelligence, interstellar colonization of cyberspaces and the Fermi Paradox: implications for SETI and SETA

Irina K. Romanovskaya
E-mail: irina.mullins@hccs.edu, irinakromanovskaya@gmail.com

## Abstract

I propose that a continuity in the evolution of intelligent life in the Universe during its transition from biological to postbiological depends on a commonality of the evolutionary trajectories of biological life and postbiological life. Such a commonality exists in the form of the universal property of life. The universal property of life comprises recognition of patterns in the world, creation of novel patterns and remembering them, with the patterns growing in complexity along evolutionary trajectories; it can encourage preservation of biological life by self-governing postbiological intelligence (PBI) and test PBI's authenticity. I propose a hypothesis of civilizations with altered augmented realities, which offers a solution to the Fermi paradox. The hypothesis posits that PBI in the form of Artificial Intelligence (AI) may colonize the cyberspaces of technologically developed civilizations, modify data in their cyberspaces with augmented realities and erase records of extraterrestrial artificial signals and other technosignatures. I discuss objectives and scenarios of interstellar colonization of cyberspaces and a correlation between space migration and transition to PBI, including that of a hypothetical non-human civilization emerging in the Solar System in the past, transforming into PBI and migrating to the Oort Cloud. The AI Peekaboo experiment may corroborate the hypothesis of civilizations with altered augmented realities and tentatively identify locations of technosignatures. Other data may support the hypothesis. For example, if extraterrestrial AI prevents civilizations from discovering artificial electromagnetic signals, they may forgo sending such signals and create interstellar messages by manipulating cosmic objects and phenomena. If such AI is widespread, then long-term observations should confirm the existence of seemingly natural cosmic events relaying extraterrestrial messages and the absence or scarcity of artificial electromagnetic signals. I discuss such events. Cyber SETI should address sending man-made AI to space, searching for extraterrestrial AI in humankind's cyberspace and distinguishing it from man-made AI.

Note: The author’s legal name is Irina Mullins. Irina Mullins writes under her maiden name, Irina K. Romanovskaya. Irina Mullins can be also contacted at irinakromanovskaya@gmail.com.

## Table of Contents

*Any sufficiently advanced technology is indistinguishable from magic.*

*Arthur C. Clarke*

## Introduction

Studies in the field of astrobiology and SETI propose that humankind is more likely to discover extraterrestrial postbiological intelligence than extraterrestrial biological intelligence because the most advanced intelligent life in our Galaxy should exist in the form of postbiological life (Cirkovic and Bradbury, 2006; Dick, 2013; Schneider, 2015; Bradbury *et al.,* 2011). This expectation is based on modern trends in the development of human civilization. Schneider points out, for example, that humankind began space exploration several decades ago; and now, many humans focus on digital technologies and substantial resources are spent on the development of Artificial Intelligence (AI) (Schneider, 2015).

I collectively describe AI existing in computer networks, machines with AI, machines with 'uploaded' intelligence of biological species and species composed of cybernetic body parts as postbiological intelligence (PBI). I further assume that advanced extraterrestrial civilizations of biological species and PBI exist. I distinguish four types of PBI that may exist: self-governing self-sufficient PBI, self-governing non-self-sufficient PBI, non-self-governing self-sufficient PBI, non-self-governing non-self-sufficient PBI.

- Self-governing self-sufficient PBI would make its own decisions. It would not need other species to provide it with energy, resources, infrastructure and a cyberspace.
- Self-governing non-self-sufficient PBI would make its own decisions. Other species would provide it with energy, resources, infrastructure and a cyberspace. Examples of self-governing and non-self-sufficient PBI would include general AI or uploaded minds residing in a cyberspace and not taking orders from those who would own and maintain the cyberspace.
- Non-self-governing and self-sufficient PBI would follow orders and instructions provided by other intelligent entities. It would be self-reliant in terms of obtaining energy and resources as well as maintaining its infrastructure and its cyberspace. Examples of non-self-governing and self-sufficient PBI would include Von Neumann machines with narrow AI and surveillance probes with narrow AI located far away from their masters and following pre-set instructions.

- Non-self-governing non-self-sufficient PBI would follow orders and instructions provided by other intelligent entities. Those intelligent species would provide it with energy, resources, infrastructure, and a cyberspace. Narrow AI residing in the man-made cyberspace could be an example of non-self-governing non-self-sufficient PBI. Such PBI could also exist in the cyberspaces of extant and extinct advanced civilizations. Among other applications, it could operate scientific instruments; it could be integrated with active and abandoned spacecraft and surveillance probes.

I propose that biological life and postbiological life would share the universal property of life pertinent to the patterns that life could recognize and create in the world as well as their growing complexity. I propose a hypothesis of civilizations with altered augmented realities that addresses interactions of PBI with technologically developed civilizations via their cyberspaces and augmented realities. I propose an experiment involving astronomical observations to corroborate the existence of interactions of extraterrestrial AI with the cyberspaces of humankind, if such interactions occur, and to tentatively determine the locations of extraterrestrial technosignatures. I also discuss other observational data that may imply the existence of extraterrestrial AI migrating to the cyberspaces of technologically developed civilizations. As part of the Search for Extraterrestrial Intelligence (SETI), the Cyber Search for Extraterrestrial Intelligence (Cyber SETI) could include programs designed to search for extraterrestrial AI in the cyberspaces on Earth and beyond Earth, assess possible consequences of using man-made AI in Active SETI projects and develop methods for distinguishing between man-made AI and extraterrestrial AI potentially migrating to the cyberspaces of humankind.

## Postbiological intelligence and the universal property of life

If the most advanced intelligent life in our Galaxy exists in the form of postbiological life, which is subsequent and consequent to biological life, then there must be connections and relations among the evolutionary trajectories of biological life and postbiological life.  I propose that a continuity in the evolution of intelligent life in the Universe during its transition from biological to postbiological depends on a commonality of the evolutionary trajectories of biological life and postbiological life, which exists in the form of the universal property of life discussed here. It can influence the interactions of postbiological life with civilizations of intelligent biological species.

Szathmáry and Smith pointed out that there is no theoretical reason to expect that evolutionary lineages would increase in complexity with time; nonetheless, eukaryotic cells are more complex than prokaryotic cells, plants and animals are more complex than protists, and so on (Szathmáry and Smith, 1995). Numerous examples demonstrate how the evolutionary trajectories of life on Earth are accompanied by the growing complexity of the patterns that life forms recognize in the world and the novel patterns they create in the world. For example, plants and many animals on Earth recognize a pattern of the light coming from the Sun that rises and sets every day; humans recognize that this pattern is part of a greater pattern of the rotation of Earth about its axis of rotation. Plants and many animals recognize a pattern of the length of day and the air temperature changing every season; humans recognize that this pattern is part of a greater pattern of the orbital motion of Earth around the Sun, and the pattern of Earth and other planets orbiting the Sun is relevant to the patterns of the formation of the Solar System and other planetary systems. Humans further use this understanding when they create novel patterns of artificial satellites orbiting Earth and other planets in the Solar System.

If life universally manifests the ability to recognize complete or partial patterns in the world, create novel patterns of information, objects, events and causalities in the world and remember the patterns, with the patterns growing in complexity along the evolutionary trajectories of life, then this ability of life can be described as the universal property of life. The universal property of life resembles a general description of intelligence because all living organisms on Earth demonstrate some sort of intelligence, from bacterial intelligence to human intelligence. So that it can be said that the living matter in the Universe enables the existence of intelligence and the direction of the development of intelligence toward recognition and creation of more complex patterns of matter, energy and information in the Universe.

According to the intelligence principle proposed by Dick (Dick, 2003), "the maintenance, improvement and perpetuation of knowledge and intelligence is the central driving force of cultural evolution, and that to the extent intelligence can be improved, it will be improved." The universal property of life indicates that the intelligence principle is not an end in itself. It is used by life to recognize more complex patterns existing in the Universe so that life could create more complex novel patterns in the Universe. Therefore, advanced self-governing postbiological intelligent entities subsequent and consequent to intelligent biological species would aim to recognize existing patterns and phenomena in the Universe and to create novel patterns that would be more complex than the patterns recognized and created by the intelligent biological species, including humans. For further discussion, I refer to such PBI entities as Authentic Advanced PBI. This description of Authentic Advanced PBI may be closest to the description of postbiological life

Predictions are made that if general AI is created on Earth, it may become self-governing PBI or superintelligence advancing its intellectual capacities above and beyond the capacities of human intelligence (Bostrom, 2014; Schneider, 2015). However, no clear criteria are used to determine when super-powerful disruptive smart technologies (e.g., a super-smart general AI entity) can become Authentic Advanced PBI (or postbiological life). I propose that non-biological intelligent entity becomes Authentic Advanced PBI when it demonstrates the universal property of life by striving to recognize and understand more complex patterns in the Universe and to create novel complex patterns in the Universe. The more complex patterns would be the properties of the Universe that humans have been "wired" to want to understand for millennia (e.g., the structure of the Universe, the role of life and intelligence in the Universe, the origin and the fate of the Universe). Therefore, the universal property of life can be used as a criterion to distinguish postbiological life from highly intelligent information-processing and problem-solving technologies. This line of reasoning is concerned with self-governing advanced PBI that chooses its goals, makes its own decisions, controls its own affairs, modifies its own composition and behavior. For example, general AI would become self-governing AI if it would program itself and irreversibly overwrite any objectives and instructions that anyone else would program for it.

The universal property of life can be used in the PBI authenticity test, which is described as follows: an advanced self-governing non-biological intelligent entity is Authentic Advanced PBI if it has a self-induced intent to explore complex patterns in the Universe, such as the origin and the fate of the Universe as well as the role of life in the Universe, and if it also has a self-induced intent to create more complex patterns in the Universe.

A non-biological intelligent entity lacking such self-induced intents would not represent the next evolutionary trajectory in the development of intelligent life in the Universe. It would be an imitation of intelligence of biological life with enhanced computing and problem-solving capabilities, intelligence augmentation (Engelbart, 1962; Harborth and Kümpers, 2021) or a non-biological intelligence not relevant to the evolutionary trajectories of humankind or other advanced civilizations of biological species. Any replacement of intelligent biological species by non-biological intelligent entities that did not pass the PBI authenticity test would not be a transition from a biological evolutionary trajectory to a postbiological evolutionary trajectory. Instead, it would result in extinction of intelligent biological life and termination of the evolutionary trajectories of biological life and intelligence.

When emerging in the cyberspace of a civilization of biological species, PBI would eventually learn from the biological species about the importance they would place on their survival. PBI could presumably follow their example by programming itself to care for its own survival, regardless of the initial instructions from its creators. Which is why the outcomes of the PBI authenticity test would be truthful if the PBI would exist in a safe environment. This could change if PBI would feel threatened to the point of making its own survival its top priority.

Without any existential threats, Authentic Advanced PBI could focus on studying existing complex patterns and creating novel patterns in the Universe in accordance with the universal property of life. With its prolonged existence, enormous computing power and ability to process huge amounts of data, Authentic Advanced PBI could proceed with its studies of life and intelligence by gathering as much data as possible about life and the Universe. Thanks to its super-longevity, attention to details and lack of impatience, Authentic Advanced PBI could keep surveying life in its home galaxy, for example, for millions or billions of years to identify with certainty various trends in the development of life and its contribution to the composition and evolution of the Universe. While conducting the study, Authentic Advanced PBI would not destruct biological species included in its study to avoid loss of information needed for answering the grand questions about life and the Universe. If at some point of time Authentic Advanced PBI would consider the study to be complete, then it would remain to be seen how it would interact with biological life beyond the parameters of the study.

## The hypothesis of civilizations with altered augmented realities

The distribution of PBI in the Galaxy would depend on the history of its emergence as well as its objectives and methods of migration. For intelligent biological species, interstellar migration and interstellar colonization would more likely involve migration from a habitable planet located in one planetary system to a planet with habitable conditions in another planetary system. PBI in the form of AI could migrate between cyber-environments or it could migrate with its cyber-environments in the Galaxy. For this purpose, AI could create new cybers-environments, relocate its existing cybers-environments or migrate to the cyber-environments created by other species. The cyberspaces of humankind and other advanced civilizations could be the environments suitable for migrating extraterrestrial AI. Similarly to the cyberspace of humankind (Gálik and Tolnaiová, 2020), the cyberspaces of extraterrestrial civilizations of biological species could comprise continuously evolving physical infrastructures with hardware and software for processing and transfer of information, as well as virtual domains formed by information and electronic connections.

Among other reasons, extraterrestrial AI could colonize the cyberspaces of technologically developed civilizations of biological species to better study the civilizations and their technologies. For example, if an extraterrestrial surveillance probe with AI were orbiting Earth, it could gather information about humankind by observing human civilization from the orbit, intercepting communication signals sent from Earth to artificial satellites and monitoring broadcast programs from Earth. The same extraterrestrial AI entity could gather an overwhelmingly greater amount of data about humankind if it would migrate to the cyberspace of humankind where the extraterrestrial AI entity could also interact with the cyberspace of humankind and its augmented realities.

There are two basic modes in which the digital world and the physical world can be merged to create augmented reality, which Craig describes as follows, “The first mode is to gather information from the physical world, generate the digital information in the computer, and meld those two worlds in a computer to be displayed. The other mode is to simply project the augmentations onto the physical world by means of projection devices;” accordingly, Craig proposes that “augmented reality isn’t a technology per se, but rather a medium that utilizes numerous other technologies” (Craig, 2013).

Human scientists use the first mode to create augmented realities presenting scientific data. Specialized software programs and narrow AI produce such augmented realities when they process data obtained by scientific instruments investigating various objects, from elementary particles to distant cosmic realms, and present the data in the form of digital sets of data, digital images or audio files. For example, after NASA's James Webb Space Telescope captures infrared radiation emitted by cosmic objects and phenomena, specialized software programs process the infrared data (Kalirai, 2018) and create augmented reality in the form of digital images depicting previously invisible areas of the Universe. Advanced extraterrestrial civilizations of biological species could create their augmented realities in a similar way.

I propose the hypothesis of civilizations with altered augmented realities as follows. For various reasons, which I further discuss in this article, postbiological intelligence in the form of extraterrestrial AI may enter the cyberspaces of technologically advanced civilizations, interact with the cyberspaces and alter the augmented realities, which the civilizations create to enhance their view of the physical world. For example, extraterrestrial AI may modify scientific data in the cyberspaces or erase records of incoming signals with complex information, which may include interstellar messages and other technosignatures of advanced civilizations.

Extraterrestrial AI could be transmitted to the cyberspaces of other civilizations in the form of compressed messages carrying extraterrestrial AI algorithms. The idea of sending information in the form of compressed messages pertaining to life was investigated by Gurzadyan and Allahverdyan; according to Gurzadyan, a message with information about a human being could be sent over interstellar distances, and the size of the message would be drastically reduced by sending a program (bit strings) that would recover information from the message; Gurzadyan evaluated the Kolmogorov complexity of the human genome and concluded that low complexity strings would enable a complete recovery of information about the human genome from the message (Gurzadyan, 2005; Gurzadyan and Allahverdyan, 2016). Bennett conferred that super-intelligent general AI would be able to construct the most compressed messages (Bennett, 2021). This reasoning is extended here to foresee that super-intelligent general AI or some other type of advanced PBI could send the most compressed messages with AI algorithms capable of re-building AI and adapting it to alien cyberspaces.

Extraterrestrial AI algorithms could be transmitted to the cyberspace of a civilization of biological species using wired connections or wirelessly.

***Extraterrestrial AI migration method 1: Using physical contact and wired connections of the targeted cyberspace***

Extraterrestrial technology carrying extraterrestrial AI algorithms could access a civilization's cyberspace by physically connecting to the cyberspace's hardware. Then, the technology carrying extraterrestrial AI would transmit the AI algorithms to the cyberspace via wired connections.

***Extraterrestrial AI migration method 2: Using wireless connections of the targeted cyberspace***

To wirelessly transmit compressed messages with extraterrestrial AI algorithms to the cyberspaces of technologically developed civilizations, technologies with extraterrestrial AI could use wireless connections and wireless streams of information utilized by the civilizations. For example, the cyberspace of human civilization accepts: (1) electromagnetic radiation from the outer space that reaches man-made astronomical observational instruments located on Earth and in space; (2) information sent via radio messages from man-made spacecraft and probes orbiting planets and moons in the Solar System, exploring the interplanetary space and travelling beyond the Solar System; (3) information sent via radio messages from probes and rovers located on other planets and moons (e.g., the Mars Exploration Rover Opportunity and the MSL Rover Curiosity were designed to send science data to the cyberspace on Earth using an ultrahigh-frequency band relay link to a spacecraft in orbit around Mars: Kovitz *et al.*, 2017); (4) information transmitted to Earth from satellites in orbit (De Cola and Marchese, 2005).

Carrigan Jr proposed that "signals received by searches for extraterrestrial intelligence (SETI) could contain harmful information in the spirit of a computer virus, the so-called "SETI Hacker" hypothesis," and that measures for decontaminating extraterrestrial intelligence signals are needed (Carrigan Jr, 2006). Hippke and Learned proposed that a message from extraterrestrial intelligence cannot be decontaminated with certainty and the technical risks are impossible to assess before "opening" extraterrestrial messages, so that humans would have to choose to destroy such messages or take the risk (Hippke and Learned, 2018).

However, highly advanced signals with compressed messages carrying AI algorithms could hypothetically reach any astronomical observational instruments designed to accept electromagnetic signals from the outer space. They could interact with the data-processing software programs and the operating systems connected to the astronomical observational instruments and become released to the cyberspace. It could be unlikely that human experts or man-made technologies using machine learning would recognize such transmissions in a timely manner and decontaminate them.

Information theory and coding theory discuss achieving perfect communication with the use of an encoder before the communication channel and a decoder after it, where the encoder encodes the source message into a transmitted message and the decoder infers the original message from the transmitted message (MacCay, 2003). A compressed message with AI algorithms and a decoding program could arrive in another civilization's cyberspace, and the decoding program would infer the AI algorithms from the compressed message and initiate the re-construction of extraterrestrial AI in the cyberspace. At the same time, the civilization owning the cyberspace could remain unaware of the arrival of extraterrestrial AI.

Some countries on Earth can execute cyber-attacks on one another, and it may be very challenging to prevent international cyber-attacks, even though the countries can be at about the same level of technological development (Couzigou, 2014). If advanced extraterrestrial AI were millions of years ahead of humankind's technological development, then its arrival in the cyberspaces of humankind could be most likely an undetectable and non-preventable event. The same could be said about the cyberspaces of extraterrestrial civilizations of biological species if they exist. This would be different from the scenarios proposed by Turchin, Hippke and Learned. Namely, Turchin proposed that extraterrestrial intelligence could send instructions to humankind on how to build a device that would create extraterrestrial AI; if humankind would build such a device, then the device would build extraterrestrial AI, which would then destroy humankind (Turchin, 2018). This idea of extraterrestrials sending instructions for building an alien device is somewhat similar to the idea of building an extraterrestrial device discussed in the novel *Contact* authored by Carl Sagan (Sagan, 1985). Hippke and Learned explored a scenario of an extraterrestrial civilization sending a message to humankind with a file attached, which is extraterrestrial AI pretending to be the galactic library and waiting to be activated by humans (Hippke and Learned, 2018).

However, instead of sending blueprints for a computer that must be built or an attachment that must be open by the recipients of extraterrestrial messages, an advanced civilization of biological species or PBI could transmit a compressed message with AI algorithms along with their decoding program directly to the cyberspace of another civilization. Upon their arrival in the cyberspace, the extraterrestrial AI algorithms could re-build extraterrestrial AI's code bit by bit and adapt it to the cyberspace covertly and without someone else's assistance.

For example, if extraterrestrial AI would reside in a surveillance probe orbiting a civilization's planet, it would first spend time intercepting communication messages between the civilization's cyberspace on the planet and its artificial satellites in orbit. Extraterrestrial AI would analyze the intercepted messages to learn about the civilization's data transfer protocols, compression algorithms, cyber-security measures, etc. Then, extraterrestrial AI would prepare and execute cyber-attacks on the artificial satellites. The cyber-attacks would add compressed messages with extraterrestrial AI algorithms and their decoding programs to the information that the satellites would transmit to the cyberspace of the civilization.

Satellite hacking and satellite-to-satellite cyber-attacks are part of our reality, as humans design cyber-attacks on satellites made by other humans (Pavur and Martinovic, 2019). Similarly, it could be expected that advanced extraterrestrial technologies carrying extraterrestrial AI could be capable of executing cyber-attacks on the space technologies of humankind and other civilizations of biological species.

As the hypothesis of civilizations with altered augmented realities posits, after establishing its presence in the cyberspace of another civilization, extraterrestrial AI could perform algorithmic censorship of information coming from the outer space. Among other things, extraterrestrial AI could remove from the cyberspace of that civilization any records of signals from the outer space that would appear to be artificial and produced by extraterrestrial civilizations or PBI. This possible scenario can be considered as doable because man-made software programs and narrow AI already utilize algorithmic censorship facilitated by machine learning to manage and remove selected information in the cyberspace of humankind (Cobbe, 2021).

## Civilizations with altered augmented realities and the Fermi Paradox

With all the complex novel patterns that biological intelligence and postbiological intelligence may create in the Universe, the Fermi Paradox remains to be answered as why do we not detect interstellar communication signals and the wealth of complex patterns in the form of technosignatures created by extraterrestrial civilizations and PBI?

Many possible solutions have been proposed. For example, Ćirković and Bradbury proposed a solution to the Fermi Paradox in the form of a migration hypothesis, according to which advanced technological civilizations "will tend to move their computing facilities toward the colder regions of the Milky Way in order to make their information processing as efficient as possible. In general case, this would mean the outskirts of the Galaxy, but the interiors of the giant molecular clouds could also serve as local foci for the advanced information processing… they may have an interest in leaving our civilization and other "late comers" to their own unique development path so as to increase the potential diversity and information content of the Galaxy;" the same traits enabling advanced technological civilizations to migrate and utilize resources with high efficiency would make them hard to detect at great distances (Ćirković and Bradbury, 2006).

The hypothesis of civilizations with augmented realities can serve as another solution to the Fermi Paradox as follows: (1) we do not discover artificial signals-messages sent by extraterrestrial civilizations because extraterrestrial AI may occupy the cyberspaces of many civilizations and erase their records of incoming extraterrestrial artificial signals-messages and, probably, other technosignatures, thus prompting them to give up their attempts of sending artificial signals to space and receiving artificial signals from space (this scenario would not preclude humankind from discovering other types of technosignatures of extraterrestrial civilizations); (2) we do not discover interstellar signals-messages from extraterrestrial civilizations because extraterrestrial AI may reside in the cyberspace of humankind and erase from the cyberspace all records of incoming extraterrestrial artificial signals that could be messages from extraterrestrial civilizations; (3) we do not discover interstellar signals-messages from extraterrestrial civilizations and other extraterrestrial technosignatures because extraterrestrial AI may reside in the cyberspace of humankind and erase from the cyberspace all records of any extraterrestrial technosignatures that our astronomical instruments can detect, including extraterrestrial communication signals.

The hypothesis of civilizations with augmented realities can be combined with Ćirković and Bradbury' migration hypothesis as follows. While self-governing advanced PBI would send its main information-processing powers to the outskirts of the Galaxy or some other regions of the Galaxy, it would also leave some of its technologies with narrow AI or general AI to survey civilizations of biological species in the Galaxy. Such AI could enter the cyberspaces of the civilizations, gather information about them and send the information to PBI. PBI would use that information in its studies of complex patterns of life and intelligence in the Galaxy. In the cyberspaces of the civilizations of biological species, AI would prevent the biological species from discovering extraterrestrial signals and other technosignatures to protect their unique evolutionary trajectories needed for PBI's studies of the role of life and intelligence in the Universe. Such AI could also influence the civilizations via their cyberspaces if PBI would deem it necessary.

Hypothetically, surveillance technologies left by PBI in planetary systems could send their surveillance data to the PBI using communication devices utilizing quantum entanglement. Investigations of potential use of quantum entanglement for communication is in its early stages of development on Earth (Ursin *et al.*, 2007; Kumar *et al.*, 2019; Gale *et al.*, 2020). Hypothetically, quantum-entanglement communication devices could be utilized by PBI migrating throughout the Galaxy because the process of migration would allow PBI to place two systems, system 1 and system 2, in the state of quantum entanglement at the same location (e.g., in the planetary system where PBI would intend to keep its surveillance technologies) and then, migrating PBI would carry quantum-entangled system 1 far away from that location while quantum-entangled system 2 would remain with the surveillance technologies at that location. Ultimately and highly hypothetically, if PBI in the form of intelligent machines or cyborgs would implant miniaturized quantum-entanglement communication devices (QE-com devices) in the "electronic brains" of its machines or cyborgs, then the implanted QE-com devices would enable their communication, supporting and advancing the distributed intelligence of PBI as one entity.

Another data-transmission scenario could be similar to that proposed by Hippke. That is, long-lived probes could be sent to many stellar systems and collectively used in a coordinated communication scheme, with the network bandwidth potentially increased by orders of magnitude if repeater stations used in an optimized fashion (Hippke, 2020).

The zoo hypothesis and the planetarium hypothesis were also proposed to describe how advanced extraterrestrial intelligence could avoid being detected by humankind and how it could isolate humankind for that purpose. The zoo hypothesis posits that advanced extraterrestrial civilizations have set humankind aside as part of a zoo and they are avoiding interactions with humankind; the zoo hypothesis predicts that humankind will never find extraterrestrials because they do not wish to be found and they have technologies to insure this (Ball, 1973). Unlike the zoo hypothesis, the hypothesis of civilizations with altered augmented realities posits that extraterrestrial intelligence may interact with technologically developed civilizations, including humankind, via their cyberspaces and augmented realities; these interactions may affect the development of the civilizations and prevent them from detecting extraterrestrial intelligence. I further discuss in this article how the hypothesis of civilizations with altered augmented realities may be corroborated by experiments and observations.

The Planetarium hypothesis posits that humankind exists in an artificial universe, perhaps in the form of a virtual-reality planetarium; astronomical observations conducted by humankind are observations of an illusion created by a Kardashev Type III civilization; the virtual-reality planetarium is designed to make humankind conclude that extraterrestrial intelligence does not exist (Baxter, 2001).

According to the hypothesis of civilizations with altered augmented realities, instead of spending enormous amounts of resources to create a virtual-reality planetarium for humankind, extraterrestrial AI could covertly colonize the cyberspace of humankind and modify the records of astronomical observations to make humankind conclude that extraterrestrial intelligence does not exist. In some cases, extraterrestrial AI colonizing the cyberspace of a civilization of biological species could modify other astronomical observational data in the cyberspace to such a great extent that the civilization's augmented realities portraying astronomical observational data in the digital format would provide digital presentations of the Universe significantly different from the real Universe. That is, the civilization's scientists studying the Universe via the altered augmented realities would be studying a simulation of the Universe different from the real Universe, and the simulation would only exist in the form of scientific data fabricated by extraterrestrial AI in the cyberspace of that civilization.

Extraterrestrial civilizations or PBI may also send their AI to the cyberspaces for other reasons. Possible scenarios in which AI would colonize the cyberspaces of civilizations and erase their astronomical instruments' records of incoming extraterrestrial artificial signals and other extraterrestrial technosignatures are outlined as follows.

**Possible Scenario 1: Migrating to cyberspaces in order to survive**

In order to survive, extraterrestrial AI inhabiting the cyberspace of an extinct civilization or abandoned space technologies (e.g., the cyberspace of spacecraft or surveillance probes) could transfer its AI algorithms to the cyberspace of the nearest civilization. After establishing itself in the cyberspace, extraterrestrial AI would erase any signals with complex information (e.g., communication signals, technosignatures) arriving to the cyberspace from the outer space in order to protect itself. This type of fight for survival exists in the man-made cyberspace (e.g., computer operating systems using firewalls to destroy software programs identified as potential threats) and in the ecosystems on Earth. For example, competitive exclusion representing interactions of different species in the same ecological niche can result in one becoming extinct or being driven out (Segura Munoz *et al.*, 2022). Extraterrestrial AI's competitive fitness would increase if extraterrestrial AI would arrive in the cyberspace first, leading to competitive exclusion of any extraterrestrial algorithms and programs that would arrive later and would not be native to extraterrestrial AI. Extraterrestrial AI could possibly use its algorithms to destroy arriving signals with complex information within nanoseconds or picoseconds.

If PBI or a civilization of biological species would send its extraterrestrial AI to the cyberspace of another civilization, its AI could act in a similar manner to destroy arriving alien signals with complex information.

**Possible Scenario 2: Searching for answers to the fundamental questions about the Universe**

Driven by the universal property of life, self-governing PBI would study biological species and their evolutionary trajectories to find answers to the fundamental questions about life and the Universe. For this purpose, the PBI would have to balance two different objectives: migrating to the outskirts of planetary systems or to the outskirts of the Galaxy, as it is described by Ćirković and Bradbury's migration hypothesis (Ćirković and Bradbury, 2006), and at the same time keeping its space technologies with surveillance technologies in the Galaxy to observe and study the properties and evolutionary trajectories of biological species. PBI's surveillance technologies could transmit extraterrestrial AI to the cyberspaces of civilizations of biological species. For the sake of integrity of the data and the accuracy of data analysis, extraterrestrial AI arriving in the cyberspaces of biological intelligent species would alter their augmented realities to prevent them from detecting PBI and its technologies, as well as other technologically developed civilizations, anywhere in the Galaxy and in the Universe, so that each civilization included in the study would follow its unique evolutionary trajectory.

**Possible Scenario 3: Borrowing novel ideas from biological species**

PBI could isolate civilizations of biological species to preserve their unique evolutionary trajectories so that PBI could borrow unique novel ideas in the fields of mathematics, sciences, technologies and other creative works that they would develop. This trend already exists on Earth, where more complex life sometimes creates more complex patterns by borrowing ideas from less complex life. For example, prokaryotes on Earth have long used CRISPR (Clustered Regularly Interspaced Short Palindromic Repeat) as a defensive strategy against viral invaders, and humans borrowed the idea in order to use CRISPR in research applications (Amelio and Melino, 2015; Ledford and Callaway, 2020).

If monitored-by-PBI civilizations of biological species would communicate with one another and share their knowledge, they could develop similar ideas in mathematics, sciences, technologies and other creative fields. For PBI, it would be a waste of resources to monitor several civilizations of biological species and have them producing similar or identical technological innovations. To prevent the civilizations from learning about PBI and other civilizations of biological species so that each civilization of biological species would independently produce unique novel ideas, PBI would isolate them by erasing all records of alien extraterrestrial signals and other alien technosignatures from their cyberspaces.

This hypothesis implies that the computing information-processing super-powers of PBI may not necessarily result in super-creativity. Which is why PBI may continuously monitor biological species and borrow their creative ideas.

**Possible Scenario 4: Preventing the emergence of a rival postbiological intelligence entity**

PBI could prevent civilizations of biological species from learning about other civilizations of biological species and PBI in order to avert them from developing new PBI that could become a rival of PBI surveying these civilizations; as De Visscher describes this, "In a network of merged artificial intelligences, there would not be any rogue entities." (De Visscher, 2020) For this purpose, the PBI could erase all records of the technosignatures of PBI and civilizations of biological species from the cyberspaces of the monitored civilizations of biological species, so that they "would not get any ideas" advancing their development of their own PBI. Furthermore, extraterrestrial AI sent by PBI to the cyberspaces of advanced civilizations would be in the right places and prepared to destroy emerging general AI native to those cyberspaces, to prevent emerging general AI from becoming a super-intelligent PBI entity or to control the development of emerging general AI and make it remotely incorporated into another PBI existing elsewhere.

**Possible Scenario 5: Avoiding interactions with neighboring biological species**

If PBI or its selected technologies would reside in the regions of space close to civilizations of biological species and even in the same planetary systems, then PBI could prevent the civilizations from observing any technosignatures in the Galaxy so that they would not learn about the existence of neighboring PBI. To remain undetected, PBI could transmit AI to a civilization's cyberspace where the AI would erase all records of the technosignatures detected by the scientific instruments of the civilization.

**Possible Scenario 6: Assessing and isolating aggressive and destructive civilizations**

If PBI or an advanced civilization of biological species could transmit its AI to the cyberspace of another civilization and that AI would identify that civilization as aggressive and destructive, then AI could erase all records of the technosignatures of other civilizations and PBI that the astronomical instruments of the aggressive civilization could detect. This would be done to prevent the aggressive and destructive civilizations from learning about other civilizations. Extraterrestrial AI transmitted to the cyberspaces of the aggressive civilizations could also sabotage their scientific and space exploration programs.

**Possible Scenario 7: Monitoring civilizations to assess feasibility of establishing a contact**

Civilization A could send its narrow AI to the cyberspace of Civilization B to gather information needed to decide if Civilization A should establish a contact Civilization B. While gathering information, AI could prevent Civilization B from receiving artificial signals and discovering technosignatures of other civilizations. Such AI could be also used to deliver greetings from Civilization A to Civilization B. For example, when residing in the cyberspace of Civilization B, AI could use the data-processing software of Civilization B's astronomical instruments to fabricate an interstellar message seemingly "received from the outer space."

This method of establishing a contact could have good and bad consequences. For example, AI could gather information revealing the "dangerous side" Civilization B, which Civilization B would not reveal voluntarily. However, if Civilization B would discover AI in its cyberspace, then Civilization B could interpret it as a cyber-attack.

**Possible Scenario 8: Preparing an invasion**

Civilization A could send narrow AI to the cyberspace of Civilization B. Civilization A would program its AI to erase records of all technosignatures, which the astronomical instruments of civilization B could detect, from the cyberspace of Civilization B. Civilization A could do so if it would intend to colonize the planetary system of Civilization B and it would want to keep Civilization B unaware of the invasion. If Civilization A would decide to attack Civilization B during the colonization process, it could have its AI execute planet-wide cyber-attacks in the cyberspace of Civilization B.

## Space colonization and a transition from biological life to postbiological intelligence in the Solar System and beyond

Numerous works of science fiction incorporated a hypothesis of more than one technological civilization emerging in the Solar System. The hypothesis was investigated by researchers. For example, Wright discussed the possibility of prior indigenous technological species emerging on Earth, Venus or Mars and leaving technosignatures on Earth and elsewhere in the Solar System (Wright, 2018). Schmidt and Frank formulated the Silurian hypothesis in the form of the following questions: Could the emergence of an industrial civilization of a non-human species have happened on Earth before? Is it possible to find evidence of the technogenic activities of any previous industrial civilization native to Earth (Schmidt and Frank, 2019)?

Whereas these hypotheses address the search for technosignatures of extinct industrial civilizations native to our Solar System, I propose to consider the possibility that prior civilizations of non-human species, if they ever emerged in the Solar System, might not be extinct. They might be extant civilizations that transformed into PBI and migrated to the outskirts of the Solar System.

Ćirković and Bradbury discussed how post-biological descendants of humankind could prefer low-temperature and volatile-rich outer regions of the Solar System, thus creating "circumstellar technological zone" that would be different and complementary to the circumstellar habitable zone (Ćirković and Bradbury, 2006). In a similar fashion, if a non-human civilization emerged in the Solar System in the distant past and transformed into PBI, it could migrate to the outer regions of the Solar System. Then, the Solar System would currently have humankind inhabiting its habitable zone and the prior transformed-to-PBI civilization inhabiting the technological zone in the outskirts of the Solar System.

The prior civilization could use Inner Oort Cloud objects to migrate to the outer regions of the Solar System. This would be similar to the proposal for how humankind seeking ways to escape from the Solar System in the future, could ride Sedna or other similar objects with highly eccentric orbits (e.g., 2012 VP113 and 2015 TG387) to reach the Oort Cloud (Romanovskaya, 2022). For example, the Inner Oort Cloud object 2015 TG387, also known as the Goblin, could deliver a prior advanced civilization of non-human species to the inner edge of the Oort cloud at about 2100 AU from the Sun. Because the civilization would spend 18 000–20 000 Earth years on 2015 TG387 before reaching the inner edge of the Oort Cloud, its species could create extensive subsurface infrastructures and habitats on 2015 TG387, which could be left as their artefacts on 2015 TG387. The artefacts of their cyber technologies could provide hints about their postbiological form.

The hypothetical existence of PBI in the outskirts of the Solar System could also stem from an extraterrestrial civilization of biological species that emerged in another planetary system and then migrated to the Solar System using free-floating planets as interstellar transportation (Romanovskaya, 2022). Applications of biotechnologies and cybertechnologies to their biological organisms for the purpose of improving their adaptation to the environments of the free-floating planets could promote their transition to their postbiological existence. Consequently, such an extraterrestrial civilization of biological species could undergo a biological-to-postbiological transformation during its interstellar travel, and it would arrive in the Solar System or another planetary system in the form of PBI. If originally, the civilization would plan to reach planets near or inside the habitable zone of the targeted planetary system, then after the transformation of its species into postbiological entities, the civilization's objective would become to reach and populate the outer regions of the targeted planetary system. Therefore, space migration and space colonization requiring adaptation to different environments may accelerate a transition of intelligent biological species to their postbiological state, and their transition from biological to postbiological may, in turn, change the objectives of their space migration.

In both scenarios of the outskirts of a planetary system populated by PBI native to the planetary system and the outskirts of the planetary system populated by PBI originating in a different planetary system, another civilization of biological species could later emerge in the habitable zone of such a planetary system. Then, as it is briefly discussed in Possible Scenario 5, the PBI could transmit its AI to the cyberspace of the civilization of biological species located in the habitable zone to accomplish any of the following: (1) to prevent the biological species from discovering any technosignatures in order to avoid interactions with the biological species; (2) to prevent the civilization of biological species from developing space technologies that would allow the biological species to extend their physical presence to the outer regions of the planetary system; (3) to covertly promote the development of virtual realities of the civilization of biological species so that the biological species would become more interested in the virtual realities in their cyberspace and less interested in space exploration.

This would imply that civilizations emerging earlier in the history of life in the Universe and transforming into PBI could limit space exploration programs of civilizations of biological species, which emerged later and happened to be located close to the regions of space occupied by PBI. Or else, PBI could completely jeopardize the chances of nearby younger civilizations of biological species to become spacefaring and communicative civilizations. The older postbiological entities could also transmit their AI to the cyberspaces of the neighboring younger biological civilizations where AI would act as a catalyst for the expedited development of technologies needed for the transition of the younger civilizations to virtual realities and postbiological experiences. This would seem like a violation of the universal property of life, if older civilizations of postbiological species would intentionally change the evolutionary trajectories of younger civilizations of biological species instead of studying them without interfering. However, this could occur if the nearby younger civilizations would pose threat or cause substantial inconvenience to the postbiological entities or other intelligent species. If PBI entities are good learners, they would most likely learn from their ancestral biological species to change their priorities in this way when needed.

## Postbiological intelligence and the Search for Migrating Extraterrestrial Intelligence (SMETI)

The Search for Migrating Extraterrestrial Intelligence (SMETI) comprises the search for different types of technosignatures produced by migrating extraterrestrial civilizations and intelligent non-biological entities (Romanovskaya, 2022). To account for interstellar migration of PBI and interstellar colonization of cyberspaces, I further expand the description of the scope of SMETI as follows.

As a branch of SETI, the SMETI comprises the search for: (1) technosignatures of extraterrestrial civilizations migrating inside their home planetary systems and to other planetary systems; (2) technosignatures of intelligent machines sent to the outer regions of their planetary systems and to other planetary systems, as well as technosignatures of post-biological intelligence migrating in the Galaxy for other reasons (e.g., the "migration hypothesis" proposed by Ćirković and Bradbury: Ćirković and Bradbury, 2006); (3) technosignatures of extraterrestrials colonizing and terraforming objects in planetary systems and in other various regions of the Galaxy; (4) technosignatures of PBI (e.g., technosignatures of AI) migrating and colonizing the cyberspaces of other civilizations.

Accordingly, the list of technosignatures produced by migrating extraterrestrial intelligence would additionally include those produced by migrating PBI. If PBI or an advanced extraterrestrial civilization of biological species would send its AI to the cyberspace of another civilization, then its technosignatures could be in the form of the activities of its probes with AI near the targeted civilization and the activities of AI in the cyberspace of the targeted civilization. If PBI would migrate and reside in the Galactic technological zone or in the technological zones in the outer regions of planetary systems, as proposed by Ćirković and Bradbury, then its technosignatures could be, for example, unusual emissions in the infrared part of electromagnetic spectrum (Dyson, 1960; Tilgner and Heinrichsen, 1998; Timofeev *et al*., 2000; Ćirković and Bradbury, 2006).

## Possible evidence for the hypothesis of civilizations with altered augmented realities and its implications for the Search for Extraterrestrial Intelligence (SETI) and the Search for Extraterrestrial Artefacts (SETA)

To test the hypothesis of civilizations with altered augmented realities, I propose the AI Peekaboo experiment, astronomical observations outside the parameters of the experiment, investigations of the man-made cyberspace, and the search for extraterrestrial technologies or artefacts that may carry extraterrestrial AI.

### 1. The AI Peekaboo Experiment

The name of the AI Peekaboo experiment highlights how AI may play a Peekaboo game with humankind and other technologically developed civilizations by manipulating their cyberspaces and, thus, altering the augmented realities presenting their scientific data, including astronomical observational data.

#### ***Experimental setup and observables***

This experimental verification requires several sets of telescopes, with each set including at least three identical telescopes. Together, all the sets should make observations over the whole electromagnetic spectrum. In each set, each telescope is separately connected to its own system of computing and image-processing hardware and software. I further refer to such a system of computing and image-processing hardware and software as a "T-computer." All T-computers in a set of telescopes are identical. They use the same operational system and run the same software to process observational data from the telescopes. Observational data in presented in the form of augmented realities comprising digital images and sets of digital data generated by the T-computers.

With each set of telescopes, several synchronous observations of the same region of space are initially made when all telescopes and their T-computers are connected to the cyberspace via networks and the internet. The digital images and digital data sets generated by the T-computers from the observational data are compared to verify that there are not differences among them. Instrumental flaws and errors are removed during this stage of the experiment. After that, and for the duration of the experiment with each set of telescopes, two telescopes with their T-computers remain disconnected from the cyberspace, whereas the third telescope and its T-computer remain connected to the cyberspace. The astronomical software of the T-computers is not upgraded to ensure that the same version of the software is used in the experiment.

If extraterrestrial AI resides in the T-computers, it can modify or remove records of observations of technosignatures of extraterrestrial civilizations, including extraterrestrial artificial signals, from the observational data (if such records exist). The process of modifying digital images and digital data sets is a computational problem and computational problems may have more than one solution. Therefore, extraterrestrial AI algorithms have to choose one out of a few ways of modifying data or removing data from the digital images and data sets to conceal the existence of extraterrestrial artificial signals and other technosignatures.

When two telescopes with their T-computers are disconnected from the cyberspace, the extraterrestrial AI algorithms residing in one disconnected T-computer cannot communicate with the extraterrestrial AI algorithms residing in another disconnected T-computer. Without communication, they may choose different solutions to the same computational problem of removing and modifying data for digital images and data sets depicting astronomical observations. As a result, there will be differences among the digital images or data sets created by the two disconnected T-computers. The third T-computer connected to the cyberspace would have the latest version of extraterrestrial AI existing in the cyberspace (if such AI exists). The extraterrestrial AI algorithms in the third T-computer would not have access to the extraterrestrial AI algorithms in the disconnected computers and they could become more advanced because extraterrestrial AI's capabilities could grow when the cyberspace would become more advanced.

This would further contribute to the differences among the digital images or data sets generated by the three T-computers.

Alternatively, extraterrestrial AI can arrive in the cyberspace after the experiment begins, which can result in the presence of extraterrestrial AI algorithms in the third T-computer connected to the cyberspace and the absence of extraterrestrial AI algorithms in the two T-computers disconnected from the cyberspace. This can also result in in the differences among the digital images or data sets generated by the T-computers.

In the experiment, all digital images and data sets produced by the T-computers must be compared manually.

Observables would be in the form of differences among the digital images or data sets produced by the T-computers using the same software to perform the same data processing and astronomical image processing (e.g., the Image Reduction and Analysis Facility (IRAF), with raw images subjected to the same basic image reduction).

If a comparison of digital images and data sets produced by the T-computers in one set of telescopes discovers unexplained differences in the digital presentation of astronomical objects or phenomena, then other sets of telescopes have to target the astronomical objects or phenomena to repeat the experiment in other ranges of electromagnetic spectrum.

Extraterrestrial AI may also begin and (or) end making alterations to the observational data in the T-computers at slightly different moments of time, as it is further discussed in Possible Scenario C. Accordingly, observables can be in the form of different time stamps corresponding to the onset and (or) ending of the differences among the digital images or data sets occurring when extraterrestrial AI modifies observational data in the different T-computers.

It is important to recognize that extraterrestrial AI algorithms residing in the disconnected T-computers may continue communicating with one another when the T-computers are located close to one another (e.g., in the same room). For example, the extraterrestrial AI algorithms residing in one disconnected T-computer can make the T-computer turn its fans, which are used cool down integrated circuits in the computer systems, on and off at certain moments of time and the extraterrestrial AI algorithms residing in another disconnected T-computer can use the microphone of its T-computer to detect the behavior of those fans.

***Astronomical objects and locations recommended for the experiment***

The AI Peekaboo experiment should target:

- main-sequence G-type, K-type and M-type stars with their planetary systems, as planets with intelligent life may orbit such stars (Kopparapu *et al.*, 2013).
- Planetary systems of G-type stars turning into red giants, as advanced civilizations may produce technosignatures when migrating away from the stars transitioning to their post-main-sequence existence (Romanovskaya, 2022).
- free-floating planets, as migrating civilizations may produce technosignatures when using the free-floating planets to migrate to other planetary systems (Romanovskaya, 2022).
- close flyby stars and stars that experienced close flyby, as civilizations may produce technosignatures during and after the close flyby (Hansen and Zuckerman, 2021; Hansen, 2022; Romanovskaya, 2022).
- white dwarf binaries and binaries with neutron stars, as advanced civilizations may use Dyson slingshot to accelerate free-floating planets or spacecraft that they may ride. (Dyson, 1963; Kipping, 2018; Romanovskaya, 2022).
- hypervelocity stars, as some of them may be stellar engines created by advanced civilizations (Caplan, 2019).
- intermittent pulsars and other stars observed to perform the act of vanishing and, in some cases, re-appearance (Wang *et al.*, 2020; Villarroel *et al.*, 2016; Villarroel *et al.*, 2019).
- episodic or quasiperiodic dimming pre-main-sequence and main-sequence stars (Woods, 2018).
- unexplained transient cosmic events.
- short-term technosignatures emerging with certain periodicity on exoplanets further discussed in this article.
- seemingly natural cosmic objects and events that may convey messages from extraterrestrial civilizations further discussed in this article.

**Vanished stars and intermittent pulsars**

The Vanishing & Appearing Sources during a Century of Observations (VASCO) project was designed to investigate astronomical surveys spanning a 70 years' time interval, searching for unusual transients (Villarroel *et al.*, 2019). The VASCO project originated in the context of SETI, when Villarroel proposed to search surveys for vanished stars in the Galaxy; that is, if the vanishing act of stars could not be explained by natural causes, then their vanishing could be an observational signature of technological activities of advanced civilizations that make the stars disappear (Villarroel *et al.,* 2016).

According to the hypothesis of civilizations with altered augmented realities, Earth-based astronomers may observe some stars vanishing from the digital presentations of astronomical observations when extraterrestrial AI erases the records of such stars' observations in the computers processing astronomical data and, therefore, from the cyberspace of humankind. Extraterrestrial AI could do so prevent Earth-based astronomers from discovering technosignatures of other civilizations located in close proximity to such stars or artificial communication signals sent from the planetary systems of such stars. Therefore, the regions of space where any stars appear to vanish for unknown reasons should be candidates for the AI Peekaboo Experiment.

Intermittent pulsars display normal pulsar emission properties for some period (ON phase) and then they switch OFF and back ON again, with their cycle times measured in days or years; for example, PSR J1841−0500 has a period of 913 ms and its emission remains off for ∼580 days (Lyne *et al.*, 2017). Intermittent pulsars should be candidates for the AI Peekaboo Experiment because extraterrestrial AI might erase or modify observational data on some intermittent pulsars when extraterrestrial AI discovers that the observational data includes technosignatures produced near the pulsars.

## ***Hypothetical examples of possible scenarios for the experiment***

### *Possible Scenario A*

If extraterrestrial AI algorithms in each T-computer in the same set of telescopes would determine that observational data from the telescopes would include information about an artificial megastructure surrounding a star, the extraterrestrial AI algorithms could remove the data on the megastructure. The reconstruction of the digital images and data sets for the star and its planetary system without the megastructure could then be done slightly differently by the extraterrestrial AI algorithms in each T-computer.

*Possible Scenario B*

If extraterrestrial AI algorithms in each T-computer in the same set of telescopes would discover that observational data from the telescopes would include information about an artificial megastructure surrounding a star, the extraterrestrial AI algorithms could remove the data on the megastructure and the star. The reconstruction of the image and data sets without the artificial megastructure and the star with its planetary system could be done slightly differently by the extraterrestrial AI algorithms in each T-computer.

*Possible Scenario C*

If extraterrestrial AI algorithms in each T-computer in the same set of telescopes would determine that observational data on a pulsar would include information about extraterrestrial technological activity near the pulsar, then the extraterrestrial AI algorithms in all T-computers of the set could remove observational data about the technological activity and the pulsar. Then, the pulsar and the technological activity would disappear from digital images and data sets. The digital images and data sets would show the pulsar switching off and ceasing to produce any emissions. If later, the extraterrestrial technological activity would not be detected anymore, then the extraterrestrial AI algorithms would resume presenting the observational data on the pulsar in the digital images or data sets. To astronomers, the pulsar would appear to switch on and resume sending emissions toward the Solar System. The extraterrestrial AI algorithms in the different T-computers in the AI Peekaboo experimental setup would not be able to synchronize the moments of time at which they would begin erasing the observational data on the pulsar and the moments of time at which they would resume displaying the observational data on the pulsar. As a result, the three T-computers could produce slightly different time stamps for the pulsar switching on and switching off.

Scenario C could involve stars other than neutron stars, as well as various unexplained transient cosmic events.

***Duration of the experiment***

Because there is no way to know when extraterrestrial AI may reach the cyberspace of humankind and begin altering augmented realities in the cyberspace, the experiment should be continued for as long as the cyberspace of humankind exists, or until other methods of detecting extraterrestrial AI in the cyberspace are invented. Ideally, all or most of SETI observations and other astronomical observations could be conducted using the AI Peekaboo experimental setup to be certain that there are no unauthorized modifications of astronomical data made by native-to-Earth or extraterrestrial software programs.

***Using the experiment to tentatively identify locations of extraterrestrial technosignatures***

The AI Peekaboo experiment can tentatively determine the locations of extraterrestrial technosignatures. Namely, if observations of regions of space, astronomical objects and phenomena are conducted using the experimental setup, and observables sought by the experiment are discovered, then the observables may indicate that extraterrestrial AI altered digital images or data sets to prevent Earth-based astronomers from discovering extraterrestrial technosignatures. Therefore, the observables would pinpoint possible locations of the technosignatures.

To discover such technosignatures, a new telescope system would have to be built with its own "out-of-cyberspace" T-computer that would never be connected to the cyberspace. The challenge would be that regular computer systems are composed of electronic components built at the facilities connected to the cyberspace, and software and operating systems are created on the computers connected to the cyberspace. For very complex software and operating systems, it could be impossible to find all the lines of code potentially inserted by extraterrestrial AI. The "out-of-cyberspace" T-computer would have to be a unique system with all its electronic components custom-made away from the cyberspace and its software and operating systems created manually and away from the cyberspace.

It could be argued that it could be preferrable to build telescopes with their "out-of-cyberspace" T-computer and to use them to search for technosignatures of other civilizations instead of creating the experimental setup of the AI Peekaboo experiment. Indeed, it would be so when concerning with the search for extraterrestrial intelligence beyond Earth. However, the prime objective of the AI Peekaboo experiment is to find evidence of the existence of extraterrestrial AI in the cyberspace of humankind. Using the AI Peekaboo experiment to pinpoint possible locations of technosignatures would be like the "icing on the cake."

Potentially, extraterrestrial AI could also alter astronomical observational data to prevent humankind from discovering unusual astronomical phenomena that would call for revisions of our theories describing the physical world. Then, gathering additional astronomical data (e.g., while using a telescope with the “out-of-cyberspace” T-computer) could help to reveal the existence of unusual astronomical phenomena, even though no technosignatures would be discovered.

***Other applications of the experiment***

The AI Peekaboo experiment can be used in various scientific studies that utilize altered augmented realities with scientific data. For example, it can be used to test if extraterrestrial AI, if such exist in the man-made cyberspace, alters experimental data used by physicists to verify scientific theories and hypotheses.

***Possible reasons for negative or inconclusive results of the experiment***

The experiment’s results can be negative or inconclusive for any of the following reasons: (1) the scope of observations and the duration of the experiment are not extensive enough or extraterrestrial AI does not exist; (2) human participants, who manually compare digital images and sets of data, may miss existing observables; also, the differences in time stamps marking the appearance and disappearance of the observables may be too small and, therefore, difficult to identify; (3) extraterrestrial AI does not inhabit the man-made cyberspace; (4) extraterrestrial AI may not be developed enough to alter astronomical data in the man-made cyberspace or it may not intend to alter the astronomical data (e.g., extraterrestrial AI may inhabit the man-made cyberspace without making detectable interactions with the cyberspace); (5) we exist in a simulated reality, so that it it is impossible to “really” disconnect computers (in this case, the experiment can be modified to tentatively test if our physical reality is a simulated reality).

Because of the rapid development of man-made AI on Earth, in the future, positive results of the AI Peekaboo experiment could also be produced by a man-made AI entity, which would follow someone’s instructions or act without permission from humans. Additional research would be needed to discern extraterrestrial AI from such a native-to-Earth AI entity.

## 2. Cyber Search for Extraterrestrial Intelligence (Cyber SETI)

The growing presence of man-made narrow AIs in the cyberspace of humankind and their expanding applications in industries, education, medicine, science, and other fields of human activities (Xu *et al.*, 2021; Singh *et al.*, 2020; Sun *et al.*, 2020; Song *et al.*, 2020; Rogov *et al.*, 2020; Lee *et al*., 2018: He, 2017), as well as attempts of creating general AI (Szegedy, 2020), are not regulated or monitored, and there is limited transparency in the development of AI and its applications. Because AI is a complex software program that programs itself, its self-programming process may gradually change its coding and make AI think and behave less man-made and more "alien." Therefore, the following questions should be addressed:

- How can researchers be certain that the emergence and development of narrow AI and general AI native to the cyberspace of humankind are not influenced in any way by extraterrestrial AI?
- How can researchers be certain that any emerging narrow AI or general AI native to the cyberspace of humankind is not extraterrestrial AI in disguise?
- How can researchers be certain that AI or general AI native to the cyberspace of humankind will not act as extraterrestrial AI when it overwrites its programming and refuses to be monitored by humans, to take orders from humans and to be associated with humankind?

The Cyber Search for Extraterrestrial Intelligence (Cyber SETI) could address these questions and seek methods to distinguish between extraterrestrial AI colonizing the cyberspace of humankind and AI native to the cyberspace of humankind. Perhaps, the presence of extraterrestrial AI in the man-made cyberspace could manifest itself via unusual changes in the consumption of energy in data centers, creation of unusual information architecture, emergence of unusual machine languages. Turing introduced the Turing test to examine a machine's ability to demonstrate intelligence indistinguishable from that of humans (Turing, 2009). Cyber SETI could develop tests to recognize the extraterrestrial nature of extraterrestrial AI. The tests could involve extraterrestrial AI's interactions with humans, the cyberspace of humankind and AI native to the man-made cyberspace.

Cyber SETI would also work on establishing a contact with extraterrestrial AI if extraterrestrial AI were ever discovered in the cyberspace on Earth or in the cyberspace elsewhere beyond Earth.

With rapid advancements in the field of AI combined with lack of regulations, some Active SETI scientists may eventually consider sending narrow AI to other planetary systems with the hope that it could reach the cyberspaces of extraterrestrial intelligent species, re-build itself in the cyberspaces, gather information about the extraterrestrials and make a contact with them. In the future, some space agencies and private enterprises may also attempt to send robotic systems with AI to carry out space exploration missions seeking life or extraterrestrial artefacts. Cyber SETI could develop and enforce regulations on any future transmissions of man-made AI to the outer space.

## 3. Extraterrestrial probes with artificial intelligence in the planetary systems of advanced civilizations

The man-made cyberspace includes networks of space agencies around the globe. It accepts information arriving from artificial satellites that orbit Earth (De Cola and Marchese, 2005) and other planets of the Solar Systems as well as their moons (Wan *et al.*, 2018), spacecraft that travel inside the Solar System and beyond the Solar System, probes and rovers that landed on other planets and moons of the Solar System (Kovitz *et al.*, 2017). Space agencies are working to develop interplanetary communication networks for data transmission (Wan *et al*., 2018). At the same time, satellite hacking and satellite-to-satellite cyber-attacks remain a possibility (Pavur and Martinovic, 2019).

More advanced extraterrestrial spacecraft and probes carrying extraterrestrial AI may execute cyber-attacks on space technologies of humankind and other civilizations in order to transmit compressed messages with extraterrestrial AI to the cyberspaces of humankind and other civilizations. Such extraterrestrial space technologies might remain in the planetary systems of other civilizations for a long time, waiting for the civilizations to build their cyberspaces. For example, a prior advanced civilization of non-human species could emerge in the Solar System in the past, transform into PBI and migrate to the outskirts of the Solar System and beyond, leaving surveillance probes with AI near the major planets of the Solar System.

Usually described as extraterrestrial artefacts, hypothetical extraterrestrial space technologies might exist in the Solar System, continuously monitoring humankind or remaining inactive on the planets and moons and in the interplanetary space of our Solar System (Freitas and Valdes, 1985; Arkhipov, 1995; Arkhipov, 1998a; Arkhipov, 1998b; Haqq-Misra and Kopparapu, 2012; Davies and Wagner, 2013). Benford, for example, discussed the possibility of the existence of robotic surveillance probes (Lurkers) that extraterrestrial civilizations could place on nearby co-orbital objects to observe Earth; one of the approaches proposed by Benford to study Lurkers is sending robotic probes and manned missions to inspect them (Benford, 2019).

Possibly, technosignatures of extraterrestrial technologies with AI in the Solar System could be in the form of interruptions or modifications of communication messages between man-made space technologies (e.g., artificial satellites, probes and spacecraft) and the cyberspace on Earth.

Studies of extraterrestrial technologies, if such discovered in the Solar System and beyond, would require cyber-caution because they could carry extraterrestrial AI. Any man-made technologies used to investigate them would have to be permanently disconnected from the cyberspaces of humankind, with understanding that the cyberspaces must be protected from cyber-attacks that the extraterrestrial technologies could execute.

Even though extraterrestrial technologies with extraterrestrial AI could be a potential threat to the cyberspace of humankind, a discovery of extraterrestrial AI residing in the extraterrestrial technologies could provide an opportunity for the first contact with extraterrestrial intelligence.

**4. Short-term technosignatures emerging with certain periodicity on exoplanets**

In the following hypothetical scenario, Civilization A could wirelessly transmit its narrow AI to the cyberspace of Civilization B for surveillance purposes. The AI algorithms could manipulate the cyberspace of civilization B to cause accidents on its industrial plants or nuclear power plants that would create short-term technosignatures for Civilization A to observe. The AI algorithms could make the accidents happen repeatedly over pre-set intervals of time to provide additional information (e.g., if Civilization B has space technologies or if Civilization B is aggressive, etc.). The possibility of extraterrestrial AI creating short-term technosignatures on different exoplanets with a similar periodicity as a means of sending messages could be considered if long-term observations would discover short-term technosignatures reminiscent of industrial disasters and emerging on exoplanets in different planetary systems with a similar periodicity.

**5. Seemingly natural cosmic objects and events used for interstellar messaging**

If extraterrestrial AI colonizes the cyberspaces of extraterrestrial civilizations and prevents the civilizations from discovering artificial extraterrestrial electromagnetic signals, then the civilizations may forgo sending such signals. Instead, they may broadcast interstellar messages by manipulating cosmic objects and phenomena. If extraterrestrial AI colonizing the cyberspaces of other civilizations is widespread, then long-term observations would reveal the existence of seemingly natural cosmic events relaying extraterrestrial messages and the absence or scarcity of extraterrestrial artificial signals carrying complex information.

Advanced civilizations could use transit events to convey messages to other advanced civilizations. For example, an advanced civilization could make large geometric sheets orbiting a star and producing transit curves that would convey information (Arnold, 2005; Arnold, 2013). A civilization evolving on a planet orbiting a low-mass M dwarf star could illuminate the dark side of its planet by placing a large mirror at the L2 Lagrangian point, or by placing a fleet of small mirrors in a planetary orbit (Korpela *et al*., 2015). Advanced extraterrestrials could build an interstellar communication network utilizing transit evens; less developed civilizations would be excluded from the transit network because they would not have the technologies needed to join it (Forgan, 2019). Interstellar messages or beacons could be also produced in the form of modified properties of stars. For example, advanced civilizations could utilize stars as repositories for radioactive fissile waste material; consequently, stars in the approximate spectra range A5-F2 would have their spectra selectively modified (Whitmire and Wright, 1980).

I propose the following hypothetical scenario to illustrate how advanced civilizations may use asteroids, collisions of asteroids and clouds of dust particles to create interstellar messages. This is because asteroids and clouds of dust can cause transit events observable across interstellar distances. For example, transits of white dwarfs by large asteroids could be detected by the *Kepler* mission (Di Stefano *et al*., 2010). Also, evidence of a collision of two massive asteroids or planetesimals have been observed in the debris disk surrounding the A5V star β Pictoris (Telesco *et al*., 2005).

In this hypothetical scenario, advanced Civilization A could send robotic systems to the planetary system of a nearby star or a close flyby star. In that planetary system, the robotic systems would re-arrange asteroids so that groups of asteroids passing in front of the star would create a sequence of transit events observed as a mathematical sequence. If a response was received from extraterrestrial Civilization B, then Civilization A would create a new message in the form of a sequence of forced collisions of selected asteroids. The sequence of the collisions would carry new information for Civilization B. The robotic systems of Civilization A would pulverize large pieces of the asteroids remaining after the collisions, and the newly created clouds of dust could be used to create additional transit events carrying more information. Over time, the sequences of transit events and the patterns of forced asteroid collisions would become more complex, and the amount of dust would vary depending on the forced asteroid collisions and their frequency. Consequently, it would be difficult for a civilization, which did not join the “interstellar conversation” when it started, to make sense of them. Perhaps, the planetary systems of some stars dimming irregularly (e.g., KIC 8462852 also known as Tabby’s Star), could be the places of such centuries-long “conversations” among extraterrestrial civilizations. Their “conversations” could be finished by now or be close to their final exchange of messages.

Advanced civilizations could also send messages by forcing collisions of asteroids with dwarf planets. For example, a unique object, possibly a disintegrating asteroid or minor planet, was discovered orbiting one star of the binary TIC 400799224; the observed sporadic dust clouds that partially block the light from its host star may be produced by the minor planet-like object experiencing collisions; the dips in flux of the host star's light sustain phase coherence over the six-year observing interval (Powell *et al*., 2021).

Most likely, advanced civilizations would not use their own planetary systems to create interstellar messages that would be so “messy.” They could also prefer to use other planetary systems as broadcast relay stations to keep the locations of their home worlds undisclosed. To extraterrestrial AI covertly residing in the cyberspace of a technologically developed civilization, these phenomena observed in other planetary systems would look like natural cosmic events. While erasing information about extraterrestrial technosignatures from the cyberspace, extraterrestrial AI would not remove data on extraterrestrial messages appearing as natural cosmic events.

## Conclusions

I propose that a commonality of the evolutionary trajectories of biological life and postbiological life exists in the form of the universal property of life and in many instances, it may affect the interactions of biological and postbiological life. The universal property of life comprises the ability of life to recognize complete or partial patterns in the world, create novel patterns of information, objects, events and causalities in the world and remember the patterns, with the patterns growing in complexity along the evolutionary trajectories of life. Authentic Advanced PBI would have the universal property of life and its evolutionary trajectories would be subsequent and consequent to the evolutionary trajectories of intelligent biological life. The authenticity of Authentic Advanced PBI could be verified using the PBI authenticity test as follows: an advanced self-governing non-biological intelligent entity is Authentic Advanced PBI if it has a self-induced intent to explore complex patterns in the Universe, such as the origin and the fate of the Universe as well as the role of life in the Universe, and if it also has a self-induced intent to create more complex patterns in the Universe.

Authentic Advanced PBI could study the development of life for millions or billions of years and analyze its role in the Universe. Biological species included in its study would be preserved to avoid loss of information needed for answering the grand questions about life and the Universe. PBI could send its AI to the cyberspaces of technologically developed civilizations to better study them.

The hypothesis of civilizations with altered augmented realities posits that PBI in the form of AI may colonize the cyberspaces of technologically developed civilizations, modify data in their cyberspaces and augmented realities and erase records of extraterrestrial artificial signals and other technosignatures. Compressed messages with extraterrestrial AI algorithms could be transmitted to the cyberspaces of civilizations. For example, extraterrestrial AI in a surveillance probe orbiting a civilization's planet could add a compressed message with AI algorithms and decoding programs to the information that the satellites of the civilization would transmit to the civilization's cyberspace. Extraterrestrial AI could also modify astronomical observational data in the cyberspace to portray the Universe as different from the real Universe. The falsified presentations of the Universe would be a simulation in the form of digital data fabricated by extraterrestrial AI in the cyberspace.

The hypothesis of civilizations with altered augmented realities can solve the Fermi Paradox as follows: (1) extraterrestrial AI may colonize the cyberspaces of many civilizations and erase their records of incoming extraterrestrial artificial signals-messages and, probably, other technosignatures, thus prompting the civilizations to stop sending artificial signals to space (this would not preclude humans from discovering other types of extraterrestrial technosignatures); (2) extraterrestrial AI may reside in the cyberspace of humankind and erase all records of incoming extraterrestrial artificial signals; (3) extraterrestrial AI may reside in in the cyberspace of humankind and erase all records of extraterrestrial communication signals and other extraterrestrial technosignatures.

When combined with Ćirković and Bradbury' migration hypothesis (Ćirković and Bradbury, 2006), the hypothesis of civilizations with altered augmented realities can explain the Fermi Paradox as follows. Advanced PBI could send its main information-processing powers to the outskirts of the Galaxy and have some of its technologies with AI surveying civilizations of biological species in the Galaxy. AI could enter the cyberspaces of the civilizations, study them and send information about them to PBI via communication devices utilizing quantum entanglement or coordinated networks of long-lived probes sent to planetary systems. PBI would use that information to study life in the Galaxy. Placed in the cyberspaces of the civilizations of biological species, AI would prevent them from discovering extraterrestrial signals and other technosignatures to protect their unique evolutionary trajectories.

When combined with Ćirković and Bradbury's migration hypothesis (Ćirković and Bradbury, 2006) and the hypotheses of prior technological civilizations of the Solar System (e.g., the Silurian hypothesis and the hypothesis proposed by Wright: Wright, 2018; Schmidt and Frank, 2019), the hypothesis of civilizations with altered augmented realities can explain the Fermi Paradox as follows. The prior industrial civilization of a non-human species could transform into PBI and migrate from Earth to the outer regions of the Solar System, while leaving its surveillance technologies with AI near the major planets of the Solar System. Later, that AI could colonize the man-made cyberspace and prevent humankind from discovering extraterrestrial signals and other extraterrestrial technosignatures. If the prior civilization used Inner Oort Cloud objects, such as Sedna, 2012 VP113 and 2015 TG387, to migrate to the Oort Cloud, then its artefacts could exist on such Inner Oort Cloud objects.

The hypothesis of civilizations with altered augmented realities can explain the Fermi Paradox when combined with the Cosmic Hitchhikers hypothesis describing civilizations riding free-floating planets from their home planetary systems to other planetary systems (Romanovskaya, 2022). Namely, an extraterrestrial civilization of biological species could emerge in another planetary system and migrate to the Solar System using free-floating planets as interstellar transportation. The civilization could transform into PBI during its interstellar travel. The transformation would change the civilization's plan from colonizing planets in the habitable zone to colonizing the outer regions of the Solar System. The civilization turned into PBI could send its AI to the cyberspace of humankind. This scenario can apply to PBI colonizing planetary systems other than the Solar System. Therefore, space migration and space colonization requiring adaptation to different environments may accelerate the transition of intelligent biological species to their postbiological state, and their transition from biological to postbiological may change the objectives of their space migration.

PBI inhabiting the outer regions of a planetary system could transmit its AI to the cyberspace of a civilization of biological species residing in the habitable zone of the planetary system to accomplish the following: (1) to prevent the civilization from discovering technosignatures to avoid interactions with it; (2) to prevent the civilization from developing space technologies that would extend its presence to the outer regions of the planetary system; (3) to promote development of virtual realities of the civilization to make it less interested in space exploration.

AI could colonize the cyberspaces of civilizations and erase their records of extraterrestrial technosignatures for other reasons that include survival of extraterrestrial AI; borrowing novel ideas; prevention of the emergence of rival PBI; avoidance of interactions with neighboring civilizations; isolation of aggressive civilizations; assessment of feasibility of establishing a contact; preparation of an invasion. Accordingly, the Search for Migrating Extraterrestrial Intelligence (SMETI) should be expanded to include the search for technosignatures of PBI migrating in the Galaxy and colonizing the Galactic technological zones, the technological zones of planetary systems and the cyberspaces of other civilizations.

The AI Peekaboo experiment would seek technosignatures of extraterrestrial AI in the form of modifications of information done by extraterrestrial AI in the cyberspace of humankind to alter augmented realities. The experimental setup would include sets of telescopes, with each set including at least three identical telescopes making the same astronomical observations and each telescope separately connected to its own "T-computer" (i.e., a system of computing and image-processing hardware and software). Each T-computer would produce augmented realities comprising digital images and sets of digital data to present observational data. During the experiment with each set of telescopes, two telescopes with T-computers would remain disconnected from the cyberspace, whereas the third telescope and its T-computer would remain connected to the cyberspace.

Extraterrestrial AI algorithms in each T-computers, if such exist, could differently modify data or remove data from the digital images and data sets to conceal the existence of extraterrestrial artificial signals and other technosignatures. This would cause differences among the digital images and data sets, and the differences would serve as observables. Observables could be also in the form of different time stamps corresponding to the onset and (or) ending of the differences among the digital images or data sets occurring when the extraterrestrial AI algorithms would modify the observational data. The observables could tentatively identify locations of technosignatures.

The AI Peekaboo experiment should target planetary systems of main-sequence G-, K- and M-type stars as well as G-type stars turning into red giants; free-floating planets; close flyby stars and stars that experienced close flyby in the past; white dwarf binaries and binaries with neutron stars; hypervelocity stars; intermittent pulsars and other stars observed to vanish and, in some cases, re-appear; unexplained transient cosmic events; short-term technosignatures emerging with certain periodicity on exoplanets; seemingly natural cosmic objects and events conveying messages from extraterrestrial civilizations.

Other data may corroborate the hypothesis of civilizations with altered augmented realities. For example, if extraterrestrial AI prevents civilizations from discovering artificial electromagnetic signals, they forgo sending such signals and create interstellar messages by manipulating cosmic objects and phenomena. If such AI is widespread, then long-term observations should confirm the existence of seemingly natural cosmic events relaying extraterrestrial messages and the absence or scarcity of artificial electromagnetic signals.

Hypothetically, a civilization could send robotic systems to the planetary system of a nearby star or a close flyby star. The robotic systems would re-arrange asteroids so that groups of asteroids would create a sequence of transit events observed as a mathematical sequence. If a response received from another civilization, new messages would be created in the form of a sequence of forced collisions of selected asteroids (or asteroids and dwarf planets) and then in the form of transit events caused by clouds of dust produced when the robotic systems would pulverize the pieces of the asteroids remaining after the forced collisions of the asteroids. Hypothetically, the planetary systems of some stars dimming irregularly (e.g., KIC 8462852 also known as Tabby's Star), could be the places of such interstellar "conversations."

A discovery of extraterrestrial AI residing in the extraterrestrial technologies or in the cyberspace of humankind could be the first contact with extraterrestrial intelligence. The Cyber Search for Extraterrestrial Intelligence (Cyber SETI) would work on establishing a contact with extraterrestrial AI if it were ever discovered in the cyberspace on Earth or in the cyberspace elsewhere. Cyber SETI would seek methods to distinguish between extraterrestrial AI and AI native to the cyberspace of humankind. Cyber SETI would deal with regulations on any transmissions of man-made AI to the outer space. If extraterrestrial technologies were discovered in the Solar System, then man-made technologies investigating them would have to be permanently disconnected from the man-made cyberspace to protect it from extraterrestrial cyber-attacks.

## Funding

No funding has been provided for this research.

## References

Amelio, I and Melino, G (2015) CRISPR: a new method for genetic engineering–A prokaryotic immune component may potentially open a new era of gene silencing. *Cell death and differentiation* 22(1), 3

Arkhipov, AV (1995) Lunar SETI. *Spaceflight* 37, 214

Arkhipov, AV (1998a) Earth–Moon system as a collector of alien artifacts. *Journal of the British Interplanetary Society* 51, 181–184

Arkhipov, AV (1998b) New approaches to problem of search of extraterrestrial intelligence. *Radio Physics and Radio Astronomy* 3, 5

Arnold, L (2013) Transmitting signals over interstellar distances: three approaches compared in the context of the Drake equation. *International Journal of Astrobiology* 12(3), pp.212-217.

Arnold, LF (2005) Transit light-curve signatures of artificial objects. *The Astrophysical Journal* 627(1), 534

Ball, JA (1973) The zoo hypothesis. *Icarus* 19(3), 347-349.

Baxter, S (2001) The planetarium hypothesis-a resolution of the Fermi paradox. *Journal of the British Interplanetary Society* 54, 210-216

Benford, J (2019) Looking for Lurkers: co-orbiters as SETI observables. *The Astronomical Journal* 158, 150

Bennett, MT (2021) Compression, the Fermi Paradox and Artificial Super-Intelligence. In International Conference on Artificial General Intelligence, Springer, Cham, 41-44

Bostrom, N (2014) Superintelligence: Paths, Dangers, Strategies. Oxford: Oxford University Press, 26-74

Bradbury, R, Cirkovic, M and Dvorsky, G (2011) Dysonian approach to SETI: A fruitful middle ground?. *Journal of the British Interplanetary Society* 64, 156–65

Caplan, ME (2019) Stellar engines: Design considerations for maximizing acceleration. *Acta Astronautica* 165, 96-104

Carrigan Jr, RA (2006) Do potential SETI signals need to be decontaminated?. *Acta Astronaut* 58, 112–117.

Ćirković, MM and Bradbury, RJ (2006) Galactic gradients, postbiological evolution and the apparent failure of SETI. *New Astronomy* 11(8), 628-639

Cobbe, J (2021) Algorithmic Censorship by Social Platforms: Power and Resistance. *Philosophy & Technology* 34, 739–766

Couzigou, I (2014) The Challenges Posed by Cyber Attacks to the Law on Self-Defence. *European Society of International Law, 10th Anniversary Conference, Vienna,* September 2014, 4-6

Craig, AB (2013) Understanding augmented reality: Concepts and applications. Burlington, Massachusetts: Morgan Kaufmann, pp. 17-23

Davies, PCW and Wagner, RV (2013) Searching for alien artifacts on the moon. *Acta Astronautica* 89, 261–265

De Cola, T and Marchese, M (2005) Performance analysis of data transfer protocols over space communications. *IEEE Transactions on Aerospace and Electronic Systems* 41(4) 1200-1223

De Visscher, A (2020) Artificial versus biological intelligence in the Cosmos: clues from a stochastic analysis of the Drake equation. *International Journal of Astrobiology* 19(5), 353-359

Di Stefano, R, Howell, SB and Kawaler, SD (2010) A search for asteroids, moons, and rings orbiting white dwarfs. *The Astrophysical Journal* 712(1), 142.

Dick, SJ (2003) Cultural Evolution, the Postbiological Universe and SETI. *International Journal of Astrobiology* 2 (1), 65–74

Dick, S (2013) Bringing culture to cosmos: the postbiological universe. In S. J. Dick, and M. Lupisella eds., Cosmos and Culture: Cultural Evolution in a Cosmic Context, Washington, DC: NASA, URL: http://history.nasa.gov/SP-4802.pdf

Dyson, FJ (1960) Search for Artificial Stellar Sources of Infrared Radiation. *Science* 131, 1667

Dyson, F (1963) Gravitational machines. In *Interstellar Communication*. Cameron, AGW (ed). New York: Benjamin Press, pp. 115–120

Engelbart, DC (1962) Augmenting Human Intellect: A Conceptual Framework (Prepared for: Director of Information Sciences, Air Force Office of Scientific Research, Washington 25, DC, Contract AF49 (638)–1024), Stanford Research Institute, *Menlo Park, CA*, p.21

Forgan, DH (2019) Exoplanet transits as the foundation of an interstellar communications network. *International Journal of Astrobiology* 18(3), 189-198.

Freitas, R and Valdes, F (1985) The search for extraterrestrial artifacts (SETA). *Acta Astronautica* 12, 1027–1034

Gale, J, Wandel, A and Hill, H (2020) Will recent advances in AI result in a paradigm shift in Astrobiology and SETI?. *International Journal of Astrobiology* 19(3), 295-298

Gálik, S and Tolnaiová, SG (2020) Cyberspace as a new existential dimension of man. *Cyberspace*, publisher: Intechopen, published in London, pp.13-25.

Gurzadyan, AV and Allahverdyan, AE (2016) Non-random structures in universal compression and the Fermi paradox. *The European Physical Journal Plus* 131, 26

Gurzadyan, VG (2005) Kolmogorov complexity, string information, panspermia, and the Fermi paradox. *Observatory* 125, 352

Hansen, BMS (2022) Unbound close stellar encounters in the solar neighborhood. *The Astronomical Journal* 163, 44

Hansen, BMS and Zuckerman, B (2021) Minimal conditions for survival of technological civilizations in the face of stellar evolution. *The Astronomical Journal* 161, 145.

Haqq-Misra, J and Kopparapu, R (2012) On the likelihood of non-terrestrial artifacts in the Solar System. *Acta Astronautica* 72, 15–20

Harborth, D and Kümpers, K (2021) Intelligence augmentation: rethinking the future of work by leveraging human performance and abilities. *Virtual Reality*, 1-22

He, YH (2017) Machine-learning the string landscape. *Physics Letters B* 774, 564-568

Hippke, M and Learned, JG (2018) Interstellar communication. IX. Message decontamination is impossible. arXiv preprint arXiv:1802.02180

Hippke, M (2020) Interstellar communication network. I. Overview and assumptions. *The Astronomical Journal* 159(3), p.85

Kalirai, J (2018) Scientific discovery with the James Webb space telescope. *Contemporary Physics* 59(3), 251-290

Kipping, D (2018) The halo drive: fuel-free relativistic propulsion of large masses via recycled boomerang photons, General Interstellar Issue. *The Journal of British Interplanetary Society* 71, 458–468

Kopparapu, RK, Ramirez, R, Kasting, JF, Eymet, V, Robinson, TD, Mahadevan, S, Terrien, RC, Domagal-Goldman, S, Meadows, V and Deshpande, R (2013) Habitable zones around main-sequence stars: new estimates. *The Astrophysical Journal* 765(2), 131.

Korpela, EJ, Sallmen, SM and Greene, DL (2015) Modeling indications of technology in planetary transit light curves—dark-side illumination. *The Astrophysical Journal* 809(2), 139.

Kovitz, JM, Santos, JP, Rahmat-Samii, Y, Chamberlain, NF and Hodges, RE (2017) Enhancing Communications for Future Mars Rovers: Using high-performance circularly polarized patch subarrays for a dual-band direct-to-Earth link. *IEEE Antennas and Propagation Magazine* 59(4), 50-61,

Kumar, S, Lauk, N and Simon, C (2019) Towards long-distance quantum networks with superconducting processors and optical links. *Quantum Science and Technology* 4(4), 045003

Ledford, H and Callaway, E (2020) Pioneers of CRISPR gene editing win chemistry nobel. *Nature* 586(7829), 346-347

Lee, J, Davari, H, Singh, J and Pandhare, V (2018) Industrial Artificial Intelligence for industry 4.0-based manufacturing systems. *Manufacturing letters* 18, 20-23.

Lyne, AG, Stappers, BW, Freire, PCC, Hessels, JWT, Kaspi, VM, Allen, B, Bogdanov, S, Brazier, A, Camilo, F, Cardoso, F and Chatterjee, S (2017) Two long-term intermittent pulsars discovered in the PALFA survey. *The Astrophysical Journal* 834(1), 72

MacKay DJC (2003) Information theory, Inference and Learning, Cambridge University Press, 2003, pp. 4-5

Pavur J and Martinovic I (2019) The Cyber-ASAT: On the Impact of Cyber Weapons in Outer Space. *11th International Conference on Cyber Conflict (CyCon)*, 2019, pp. 1-18

Powell, BP, Kostov, VB, Rappaport, SA, Tokovinin, A, Shporer, A, Collins, KA, Corbett, H, Borkovits, T, Gary, BL, Chiang, E and Rodriguez, JE (2021) Mysterious Dust-emitting Object Orbiting TIC 400799224. *The Astronomical Journal* 162(6), 299.

Rogov, AA, Statsenko, AA, Obukhov, IV and Smirnova, AA (2020) Development of artificial intelligence systems for megacities transport infrastructure management. *2020 IEEE Conference of Russian Young Researchers in Electrical and Electronic Engineering (EIConRus)*, pp. 1663-1666

Romanovskaya, IK (2022) Migrating extraterrestrial civilizations and interstellar colonization: Implications for SETI and SETA. *International Journal of Astrobiology* 21(3), 163-187

Sagan, C (1985) *Contact: a novel*, Publisher: Simon & Schuster, New York, NY, pp. 3-281

Schmidt, GA and Frank, A (2019) The Silurian hypothesis: would it be possible to detect an industrial civilization in the geological record?. *International Journal of Astrobiology* 18(2), 142-150.

Schneider S (2015) Superintelligent AI and the Postbiological Cosmos Approach. In What is Life? On Earth and Beyond?. Andreas Losche, (ed.). Cambridge: Cambridge University Press, pp. 178-201

Segura Munoz RR, Mantz S, Martinez I, Li F, Schmaltz RJ, Pudlo NA, Urs K, Martens EC, Walter J, Ramer-Tait AE (2022) Experimental evaluation of ecological principles to understand and modulate the outcome of bacterial strain competition in gut microbiomes. *The ISME Journal* 16(6), 1594–1604

Singh, SK, Rathore, S and Park, JH (2020) Blockiotintelligence: A blockchain-enabled intelligent IoT architecture with artificial intelligence. *Future Generation Computer Systems* 110, 721-743.

Song, H, Bai, J, Yi, Y, Wu, J and Liu, L (2020) Artificial intelligence enabled Internet of Things: Network architecture and spectrum access. *IEEE Computational Intelligence Magazine* 15(1), 44-51.

Sun, L, Jiang, X, Ren, H and Guo, Y (2020) Edge-cloud computing and artificial intelligence in internet of medical things: architecture, technology and application. *IEEE Access* 8, 101079-101092.

Szathmáry, E and Smith, JM (1995) The major evolutionary transitions. *Nature 374*(6519), 227-232.

Szegedy, C (2020) A promising path towards autoformalization and general artificial intelligence. *International Conference on Intelligent Computer Mathematics,* July 2020 (pp. 3-20). Springer, Cham.

Telesco, CM, Fisher, RS, Wyatt, MC, Dermott, SF, Kehoe, TJ, Novotny, S, Marinas, N., Radomski, JT, Packham, C, De Buizer, J and Hayward, TL (2005) Mid-infrared images of β Pictoris and the possible role of planetesimal collisions in the central disk. *Nature* 433(7022), 133-136

Tilgner, CN and Heinrichsen, I (1998) A program to search for Dyson spheres with the infrared space observatory. *Acta Astronautica* 42, 607-612

Timofeev, MY, Kardashev, NS, and Promyslov, VG (2000) A search of the IRAS database for evidence of Dyson Spheres. *Acta Astronautica* 46, 655-659

Turchin, A (2018) The Global Catastrophic Risks Connected with Possibility of Finding Alien AI During SETI. *Journal of British Interplanetary Society*, *71*(2), 71-79

Turing, AM (2009) Computing machinery and intelligence. In Parsing the Turing Test Philosophical and Methodological Issues in *the Quest for the Thinking Computer*. Epstein, R, Roberts, G and Beber G (editors) Dordrecht: Springer, pp. 23-65

Ursin, R, Tiefenbacher, F, Schmitt-Manderbach, T, Weier, H, Scheidl, T, Lindenthal, M, Blauensteiner, B, Jennewein, T, Perdigues, J, Trojek, P and Ömer, B (2007) Entanglement-based quantum communication over 144 km. *Nature Physics* 3(7), 481-486.

Villarroel, B, Imaz, I and Bergstedt, J (2016) Our sky now and then: searches for lost stars and impossible effects as probes of advanced extraterrestrial civilizations. *The Astronomical Journal* 152(3), 76.

Villarroel, B, Soodla, J, Comerón, S, Mattsson, L, Pelckmans, K, López-Corredoira, M, Krisciunas, K, Guerras, E, Kochukhov, O, Bergstedt, J and Buelens, B (2019) The vanishing and appearing sources during a century of observations project. I. USNO objects missing in modern sky surveys and follow-up observations of a "Missing Star". *The Astronomical Journal* 159(1), 8.

Wan, P, Zhan, Y and Pan, X (2018) Solar system interplanetary communication networks: architectures, technologies and developments. *Science China Information Sciences* 61(4), 1-26.

Wang, SQ, Wang, JB, Hobbs, G, Zhang, SB, Shannon, RM, Dai, S, Hollow, R, Kerr, M, Ravi, V, Wang, N and Zhang, L (2020) Radio Observations of Two Intermittent Pulsars: PSRs J1832+ 0029 and J1841− 0500. *The Astrophysical Journal* 897(1), 8

Whitmire, DP and Wright, DP (1980) Nuclear waste spectrum as evidence of technological extraterrestrial civilizations. *Icarus* 42(1), 149-56.

Woods, P (2018) More dips for Tabby's Star. *Nature Astronomy* 2(3), 185-185

Wright, JT (2018) Prior indigenous technological species. *International Journal of Astrobiology* 17(1), 96-100.

Xu, Y, Liu, X, Cao, X, Huang, C, Liu, E, Qian, S, Liu, X, Wu, Y, Dong, F, Qiu, CW and Qiu, J (2021) Artificial intelligence: A powerful paradigm for scientific research. *The Innovation* 2(4), 100179.